# Enhanced weathering in the U.S. Corn Belt delivers carbon removal with agronomic benefits

David J. Beerling[1*], Dimitar Z. Epihov[1], Ilsa B. Kantola[2], Michael D. Masters[2], Tom Reershemius[3], Noah J. Planavsky[3], Christopher T. Reinhard[4], Jacob S. Jordan[5], Sarah J. Thorne[1], James Weber[1], Maria Val Martin[1], Robert P. Freckleton[1], Sue E. Hartley[1], Rachael H. James[6], Christopher R. Pearce[7], Evan H. DeLucia[2] & Steven A. Banwart[8,9]

[1]Levehulme Centre for Climate Change Mitigation, School of Biosciences, University of Sheffield, Sheffield, UK

[2]Institute for Sustainability, Energy, and Environment, University of Illinois at Urbana-Champaign, IL, USA

[3]Department of Earth and Planetary Sciences, Yale University, New Haven, CT, USA

[4]School of Earth and Atmospheric Sciences, Georgia Institute of Technology, Atlanta, USA

[5]Porecast Research LLC, Lawrence, KS, USA

[6]School of Ocean and Earth Science, National Oceanography Centre Southampton, University of Southampton, Southampton, UK.

[7]National Oceanography Centre, Southampton, UK

[8]Global Food and Environment Institute, University of Leeds, Leeds, UK

[9]School of Earth and Environment, University of Leeds, Leeds, UK

*To whom ccorrespondence may be addressed. Email: d.j.beerling@sheffield.ac.uk

Author Contributions: Conceptualization: DJB, EHD, SAB, DZE, TR, NJP, CTR. Methodology: DZE, DJB, IBK, MDM, EHD, TR, NJP, CTR, CB, SJT, SHE, JSJ. Funding acquisition: DJB, NJP. Writing – original draft: DJB, DZE, SAB, NJP, CTR. Writing – review & editing: all authors.

Competing Interest Statement D.J.B. has a minority equity stake in Future Forest/Undo, remaining authors declare that they have no competing interests.

**This PDF file includes:**

    Main Text
    Figures 1 to 4

**Abstract**

**Enhanced weathering (EW) with crushed basalt on farmlands is a promising scalable atmospheric carbon dioxide removal strategy that urgently requires performance assessment with commercial farming practices. Our large-scale replicated EW field trial in the heart of the U.S. Corn Belt shows cumulative time-integrated carbon sequestration of 15.4 ± 4.1 t $CO_2$ $ha^{-1}$ over four years, with additional emissions mitigation of ~0.1 - 0.4 t $CO_{2,e}$ $ha^{-1}$ $yr^{-1}$ for soil nitrous oxide, a potent long-lived greenhouse gas. Maize and soybean yields increased 12-16% with EW following improved soil fertility, decreased soil acidification, and upregulation of root nutrient transport genes. Our findings suggest that widespread adoption of EW across farming sectors has the potential to contribute significantly to net-zero greenhouse gas emissions goals and global food and soil security.**

agricultural production | carbon removal | enhanced weathering | soil geochemistry

**Significance**

Safe, scalable atmospheric carbon dioxide removal (CDR) strategies are required for addressing the current climate emergency alongside stringent emissions reductions from industry. Enhanced weathering (EW) is a CDR strategy that involves amending farmland soils crushed rock, typically basalt, an abundant natural volcanic rock. Our results from the first long-term, large-scale EW field trial in the U.S. corn belt demonstrate reproducible carbon removal on farm fields, alongside increased soil fertility and crop yield. Our findings highlight the enormous potential for utilizing millions of hectares of farmland in the U.S. to capture carbon with EW while simultaneously improving food and soil security.

**Introduction**

Large-scale deployment of atmospheric carbon dioxide removal (CDR) strategies alongside emissions reductions will be essential for limiting future climate change caused by anthropogenic emission of $CO_2$ and other greenhouse gases (*1*). Terrestrial enhanced weathering (EW), the amendment of cropland soils with crushed silicate rocks, such as basalt, is a promising CDR strategy (*2*),(*3*),(*4*),(*5*). Purposeful EW accelerates dissolution of rock minerals to release base cations and convert atmospheric $CO_2$ into bicarbonate ions ($HCO_3^-$) which are ultimately stored in the ocean on >10,000 year timescales(*2*),(*3*),(*4*). Biogeochemical modeling suggests that deployment of EW with basalt across major agricultural regions worldwide could sequester up to two billion metric tons of $CO_2$ annually, after accounting for operational carbon emissions (mining, grinding, transport, and spreading of rock dust on fields) (*4*). In contrast to other CDR strategies, EW can improve food and soil security (*3*),(*6*) and reduce ocean acidification (*7*),(*8*),(*9*). Importantly, EW utilizes existing technology and infrastructure making it a rapidly scalable CDR option for assisting with national net-zero greenhouse gas emission plans (*10*). However, there is an urgent need to measure the rates and efficiency of EW at the farm scale with basalt in key agricultural regions.

The U.S. Corn Belt represents 70 million hectares of intensively managed agricultural land cultivating major fuel, food, and feed row crops. The region is a major contributor U.S. agricultural nitrous oxide ($N_2O$) fluxes and a key target region for quantifying the contribution of modified agricultural practices to U.S. plans for balancing the carbon budget by 2050 (e.g., *11*). We report the first findings from replicated EW field trials undertaken over four years from an experimental farm in the heart of the Corn Belt (Fig. 1; 40°30' N, 88°11'W). In our field study, we applied crushed basalt on an annual basis for four years (at a rate of 50 tons per hectare) and compared the results to control plots, in all cases using standard practices in the region for mineral nitrogen fertilization (**fig. S1**). Our farm-scale trial demonstrates the practicality of this technology in agronomy by addressing the real-world performance of EW, from carbon removal to the co-benefits for crop production and soils, across a maize (*Zea mays* L.) and soybean (*Glycine max* L.) rotation (**Fig 1A**).

We assessed the extent of EW by quantifying the mass loss of major base cations ($Ca^{2+}$, $Mg^{2+}$ and $Na^+$) from the added basalt grains by weathering, relative to the pre-treatment soils. Our approach builds from decades of tracking natural weathering rates in soils using mass balance (e.g., *14)*. We exploit a recently developed method that uses isotope dilution mass spectrometry to achieve the required analytical precision for quantifying basalt weathering rates in soils (*13*). The technique reduces analytical uncertainty of measurements by spiking soil samples with fixed amounts of base cation and titanium (Ti) isotopes ($Ca^{42/44}$, $Mg^{24/26}$, $Ti^{48/49}$). Normalization to the variable amount of basalt present in each soil sample is based on the concentration of Ti, an element enriched in the applied basalt feedstock relative to bulk soil, and which displays immobile behavior during weathering due to its low solubility (*14*). Attribution of atmospheric CDR to EW is calculated by measurement of the base cation loss from basalt and the stoichiometry of the mineral weathering reactions, following correction for strong acid weathering due to nitrogen fertilizer additions (*5*).

Conventional approaches to estimating EW and CDR are based on detailed geochemical analyses of catchment drainage waters (*15*), although conductivity has also been suggested as a proxy for cation release and alkalinity generation (*16*). Our mass balance approach is a necessary advance for detecting EW and overcoming fundamental challenges of such approaches in Midwestern croplands like the Energy Farm. These include high background alkalinity fluxes, hydrology affected by tile drainage, and uncertainty in the weathering cation mass balance due to temporary retention of weathered cations on soil exchangeable sites prior to discharge (*17-19*). Additionally, large-scale farms routinely analyse surface soils for integrated soil and nutrient management. Soil-based approaches for robust monitoring, reporting and verification (MRV) of CDR could, in theory, be readily implemented at scale on working farms.

Here, we show *in situ* soil EW rates and CDR estimation from our large-scale field trial over 4 years. We also assess how EW affects key soil health metrics over time including pH, nitrogen availability, release of phosphorus and potassium, and the pH-dependent availability of other nutrients important for healthy crops, including molybdenum and silicon. We report maize and soybean yield responses to EW over multiple years, compare the response to a traditional liming treatment, and assess root transcriptional responses of both crops to provide mechanistic insights explaining observed yield responses. Our study represents the first long-term, large-scale EW field trial with basalt on a working experimental farm, integrating evidence across geochemistry, soil science, and molecular genetics, to build a comprehensive picture of the operational performance of EW for the major agronomic ecosystem in the Midwest.

## Results and Discussion

### EW and Carbon Dioxide Removal

Based on our *in-situ* estimate of basalt weathering with EW in a conventional corn-soy rotation under field conditions, we calculate a cumulative CDR trend with a consistent increase from 4.3 t $CO_2$ ha$^{-1}$ in the first year of treatment to ~15.4 t $CO_2$ ha$^{-1}$ after 4 annual rock dust applications (**Fig. 1B**; **figs. S2-S8**). This represents a time-integrated CDR curve for our large-scale replicated EW trial, based on cation loss from the basalt grains, i.e., the potential CDR resulting from export of cations from the uppermost 10 cm of the soil profile to which ground basalt is applied. The lower soil layers showed no detectable weathering during the trial. Despite sample-to-sample variability in background soil composition, and corresponding estimated CDR, mean CDR rates are robust and similar across application years (**Fig. 1C, fig. S9**). These rates were achieved with crushed meta-basalt material characterized by a grain size *p*80 of 267 µm (defined as 80% of the particles having a diameter less than or equal to this specified particle size), with a low proportion (~11% mass) of fast-weathering minerals (*20*). Calculated initial CDR of ~4 t $CO_2$ ha$^{-1}$ per 50 t rock ha$^{-1}$ compares to a theoretical maximum CDR potential of 17.9 t $CO_2$ ha$^{-1}$ per 50 t rock ha$^{-1}$, based on the cation content of the material (see 14 and references therein, and see Datafile 1 for feedstock composition). The estimated CDR rates are comparable to mesocosm experiments with basalt addition to acidic (pH ~6.5) agricultural soil (*21*) and are notably higher than rates of soil organic carbon sequestration linked to shifts in agricultural practices (e.g., *22*). Optimization of rock dust application rates, application timing, use of finer grained material, and selection of primary basalts with faster weathering mineralogies than the by-product meta-basalt used here, offer scope for several-fold higher CDR rates (*20*).

We found no detectable increases in total inorganic carbon for a representative subset of soil samples (across blocks and sampling dates, including control and treated samples), with all measurements being below detection limit of 0.1% (Eltra C/S Analyzer). Therefore, we assume that pedogenic carbonate formation in topsoil did not constitute a significant sink for carbonate alkalinity. This is also in agreement with findings from mesocosm trials with basalt amended acidic agricultural soil (*21*).

Our approach using solid-soil analyses has the advantage of providing a time-integrated quantification of the loss of reactive mineral cations in forward weathering reactions that drive carbon sequestration. Although this avoids the need to trace and measure soluble EW CDR reaction products (alkalinity, cations, and dissolved inorganic carbon), the approach is subject to several limitations. Foremost, this method will not account for any carbon leakage in surface waters and the oceans. Further, this approach does not offer any insights into the time required for base cations to move from the soil column into surface waters, which will exceed the soil water residence time because of differential sorption onto reactive mineral phases and organic surfaces (*17-19*).

Nonetheless, the reproducible, empirical quantification of tons of CDR per hectare per year (**Fig. 1B**) provides a much-needed proof-of-concept for EW with basalt. It supports model assessment of EW and CDR capacity on farmland with basalts of differing mineral chemistry and repeated annual applications of crushed rock over consecutive years (*4*),(*5*). By demonstrating attribution of $CO_2$ removal under field conditions, we advance the feasibility of EW for the agroecosystem that is most central to U.S. agricultural production and global food security. Our results also represent an initial step to develop robust MRV of CDR through EW with basalt on farmland. However, further work is required to assess the fate of captured carbon during transport by rivers (*23*) and its release in the costal oceans (*9*) as important steps along the pathway to marine sequestration on a timescale of more than 10,000 years.

## Soil nitrous oxide emissions

Nitrous oxide ($N_2O$) is a powerful non-$CO_2$ greenhouse gas [20-yr global warming potential 273 time higher than $CO_2$ (*24*)], emitted primarily from croplands, that contributes to both climate warming and stratospheric ozone depletion (*25, 26*). In three of the four years of EW treatment, we observed reductions in cumulative soil $N_2O$ fluxes of between 14% and 41% from nitrogen-fertilized maize and unfertilized soybean plots (2017-2020) (*27*) (**Fig. 1C; fig. S10**). The mitigation of $N_2O$ emissions from basalt amended soils observed in our EW field trials is consistent with observations from mesocosm EW trials using nitrogen fertilized corn (*28*). This is a notable additional benefit of EW given that maize/soybean production in the U.S. Corn Belt accounts for up to a third of North American $N_2O$ emissions (**Fig 1D**). In contrast, other land-based CDR strategies (e.g. soil organic carbon sequestration) often increase soil $N_2O$ emissions, offsetting carbon sequestration (e.g. (*22*,*29*)).

Converted to $CO_2$ equivalents ($CO_{2,e}$), soil $N_2O$ reductions with EW improved the overall greenhouse gas (GHG) removal budgets, mitigating emissions by between 0.1 and 0.4 t ha$^{-1}$ yr$^{-1}$ (**Fig. 1E**). This compares favorably with technology-led practices to abate agricultural $N_2O$ emissions (e.g., enhanced-efficiency fertilizers, drip irrigation, and biochar) and optimization of fertilizer application rates (*29*),(*30*), many of which have additional specific costs of implementation (e.g., nitrification inhibitors ~28$/ha; (*30*)). EW practices deployed on U.S. Midwestern soils may therefore offer an effective and cost-competitive soil $N_2O$ mitigation strategy.

**Soil fertility**

In response to EW, soil pH increased significantly ($P < 0.001$) in the surface layer (0-10 cm) and deeper in the profile (10-30 cm), thereby preventing soil acidification that occurs regularly from nitrogen fertilization, as seen in control plot soils (**Fig. 2A**). This highlights the ability of basalt to potentially replace agricultural limestone application, which is commonly used to manage levels of soil acidity that often limit yields throughout the Corn Belt (*31*), while simultaneously capturing carbon and lowering other GHG emissions (**Fig. 1**). This replacement offers a substantial benefit, given soil pH regulation with limestone in the U.S. alone is considered to emit millions of tons of $CO_2$ annually (*31*). EW treated soils underwent significant ($P < 0.001$) increases in the base cation saturation of bulk soil compared to control soils, as $Ca^{2+}$ and $Mg^{2+}$ ions released from basalt weathering (**fig. S11**) replaced exchangeable acidity (**Fig. 2B**) over time. Soil pH buffering is evident in the near-neutral range in the treated plots, consistent with the protonation of soil organic and mineral surfaces (*32*). This offers evidence on the time scale of the trial that pH increase is not necessarily a limiting factor for future repeat applications of basalt.

Improved soil fertility following mineral nutrient release, and increased nutrient availability with the rise in pH as basalt weathers, are important potential co-benefits of EW alongside reversal of soil acidification (*3*),(*6*). We focus on nitrogen (N), phosphorus (P), and potassium (K), the primary nutrients supplied by expensive chemical fertilizers to lift Corn Belt yields, and molybdenum (Mo) and silicon (Si) as key subsidiary nutrients (*33*),(*34*) essential for healthy crops and yields (**Fig. 2**). We found significant ($P < 0.01$) increases in the total grain biomass N of maize and soybean in response to EW treatments (2017-2020). Increased grain N stocks in maize with EW arise largely from greater remobilization from vegetative biomass N-pools (**fig.**

S12), leading to increased ($P < 0.05$) nitrogen use efficiency [NUE; ($N_{grain(exported)}$–$N_{mineralized}$)/$N_{fertilizer}$] (**fig. S12**) (**Fig. 2 C,D**). In contrast, increased N-supply in soybean grain results from increased soil uptake and $N_2$ fixation with EW (**fig. S13, S14**). Molybdenum (Mo) is a key cofactor for nitrate reductase in biomass, the enzyme catalyzing conversion of nitrate to nitrite, which plants ultimately convert to ammonium for amino acid biosynthesis, and for symbiotic $N_2$ fixation (*35*). We show that the increased soil pH driven by EW (**Fig. 2A**) facilitates mobilization of molybdate anions in soil porewaters (**fig. S15**) in agreement with established desorption kinetics (*37*), increasing biomass Mo stocks in soybean and maize (**fig. S16)** and improving nitrogen assimilation and NUE.

In addition to N, the total mass of P and K in grain increased significantly ($P < 0.01$) with EW (**Fig. 2 E,F**). Increased P uptake into grain biomass of maize and soybean matches EW release of P from basalt (**figs. S17**). Increased total maize grain K is attributable to reallocation of K pools from vegetative biomass to the grain, whereas for soybean with relatively lower biomass K stocks, supply is matched by release from slow weathering of K-bearing silicates (**fig. S12**). Annual application of crushed basalt (50 t ha$^{-1}$) released an average of at least 7 kg P ha$^{-1}$ yr$^{-1}$ and 23 kg K ha$^{-1}$ yr$^{-1}$ by EW over four years (**fig. S17**). This highlights the potential for EW practices to partially substitute for expensive P and K fertilizers (urea phosphate, \$890 t$^{-1}$; diammonium phosphate (DAP), \$938 t$^{-1}$, potash \$862 t$^{-1}$) (*38*), with important economic and environmental savings for farmers.

Silicon (Si) is beneficial for crop productivity and resistance to stress (*36*),(*39*). Accessibility of biologically available Si declines with a rise in pH as it becomes increasingly retained on mineral surfaces (*40*). There was no significant ($P > 0.05$) change in peak biomass Si of maize and soybean in response to EW over 4 years, despite decreasing Si availability with a rise in soil pH (**Fig. 2A**), (**figs. S18-S19**). This indicates that Si release by basalt weathering compensated for reduced availability as soil pH increased. Subsequent pH-dependent soil analyses of adsorbed silica species showed an immobilized reservoir of Si released via EW that was readily mobilized with a fall in soil pH (**fig. S19**). Thus, EW can redress depletion of biologically available Si pools caused by removal of biomass as part of current U.S. farming practices which can limit yields (*36*), highlighting another advantage of EW relative to liming.

**Crop root transcriptional responses to EW**

We quantified soybean and maize root gene transcription changes in our field trials to gain mechanistic insight into the regulation of nutrient (N, P and K) uptake responses to EW during grain filling. Transcriptional profiling of the resulting 47 high-quality root RNA-Seq libraries (**fig. S20**), focusing on inorganic ion transporter genes, showed a significant proportion of transporters for ammonium and nitrate, phosphate, and potassium in soybean roots responded to basalt (**Fig. 3A**). Up-regulation of transporter genes, consistent with the specific requirements of soybean for increased soil uptake of N, P, and K during grain filling (**fig. S12**) dominate changes in root gene expression. Maize roots showed the strongest upregulation response in the group of phosphate transporter genes with EW, consistent with its requirement for soil P during grain filling (**fig. S12**), and down regulation of K and N-transporter genes, as expected with reallocation of K and N pools from vegetative biomass to grain (**Fig. 3B**). Additionally, genes expressing acid phosphatases and phytases were significantly upregulated with EW in both soybean and maize. These responses are consistent with increasing P release by enzymatic breakdown of organic matter (*41*) to assist in P uptake during grain filling. Symbiotic mycorrhizal fungal-to-root transporters of P and N (*42*),(*43*),(*44*) were upregulated in soybean, but not in maize, where high fertilizer applications lead to loss of mycorrhizal fungal partners.

Root transcriptome analyses further support the role of improved Mo availability in enhancing soybean N nutrition (*35*) with genes involved in Mo co-factor biosynthesis upregulated in response to EW, together with the genes involved in nitrate assimilation including nitrate and nitrite reductases (**Fig. 3C**). Transcriptomes showed upregulation of genes involved in conversion of root-acquired inorganic N to organic forms (pathways I to VIII) and long-distance transport genes exporting root N to aboveground soybean biomass (pathways IX and X) (*45*) (**Fig. 3C**). However, in maize where N remobilization from vegetative biomass pools was sufficient, we saw no upregulation of this suite of genes. Overall, transcriptional reprogramming of root nutrient transporters in both crops balanced grain biomass requirements in response to EW and provided genomic evidence independently supporting the observed changes in plant nutrient budgets.

**Crop yield responses**

Grain yields increased significantly with EW for maize (a $C_4$ photosynthetic crop) by 12% (2020) and soybean (a $C_3$ photosynthetic, $N_2$-fixing crop) by 16% (2019) (**Fig. 4**). Yield increases for soybean with EW are comparable to those for bioengineered soybean (*46*) and soybean grown under well-watered conditions with elevated $CO_2$; well-watered maize in contrast to EW shows no yield enhancement with elevated $CO_2$ (*47*). Our yield enhancements are broadly consistent with those from EW mesocosm trials using crushed basalt (*21*) and EW field plots and experiments using wollastonite (*48*). We attribute yield gains in both crops to the combined effects of increased nutrient supply as basalt minerals dissolve, increased nutrient availability with rising soil pH **(Fig 2)**, and greater root uptake of mineral elements (**Fig. 3**). This is supported by the lack of a significant maize yield response to liming in 2020 (**Fig. 4**), suggesting basalt provided inorganic nutrients in addition to raising soil pH. Significant yield increases in maize were only observed in those years of the rotation following soybean, suggesting an EW-nitrogen interaction (*49*). Extrapolated across the Corn Belt region, EW-related yield increases translate to ~$7-11 billion for maize, and $10 billion for soybean at current prices, indicating EW deployment could deliver substantial economic benefits for U.S. agriculture through yield increases alone.

Yield enhancements with EW were achieved with significantly ($P < 0.05$) increased key micro- and macro-nutrient concentrations (including potassium, magnesium, manganese, phosphorus, and zinc), thus improving or maintaining crop nutritional status (**fig. S21**). We observed no significant increase in the content of trace metals in grains of maize or soybean (**fig. S22**), soil pore water or soil exchangeable pools after four annual EW treatments relative to controls (**figs. S23-S24**). Molybdenum, essential for plant N metabolism increased approximately four-fold with EW over this time and copper, an essential micronutrient for photosynthesis at these levels, increased around 35%. Overall, based on 2000 measurements of twelve trace metals at two soil depths (0-10 cm and 10-30 cm) over four years of repeat basalt application, these data help alleviate concerns over possible accumulation of bioavailable metals in soils with EW (*50*).

**Conclusion**

We show with a large-scale replicated field trial over 4 years that EW drives significant CDR and mitigation of soil $N_2O$ emissions under field conditions across a conventional corn-soy rotation in the U.S. Corn Belt. Our field study demonstrates the use of *in situ* soil EW measurements that could potentially form part of a robust MRV toolkit for diagnosing rates of basalt weathering and time-integrated CDR. Such empirical quantification of $CO_2$ removal rates in the field is an essential prerequisite for MRV to facilitate wide-scale adoption of EW. Furthermore, we quantify major agronomic benefits of EW for this dominant agroecosystem, including increased crop yields (maize and soybean) via improved soil fertility and changes in root gene expressions, without adverse environmental impacts for plants and soils. Collectively, our evidence supports EW with basalt as a promising strategic management option for atmospheric CDR, deployable with existing agricultural practices and equipment. These findings highlight the capacity of EW to simultaneously augment food and soil security while delivering CDR and generating revenue for critical agricultural regions.

**Materials and Methods**

**Energy Farm, site, EW operations and field collection**

Research was conducted at the University of Illinois Energy Farm (40 3'46" N, 88 11'46" W), south of Urbana, Illinois, in 2016-2020. Energy Farm research has broadly focused on perennial bioenergy agronomy, land conversion from annual to perennial, and other ways to store belowground C in an area boasting over 100 years of maize and soybean production. Mean annual temperature is 11°C and mean annual precipitation is 1051 mm, evenly distributed throughout the year. Average N deposition measured at Bondville, Illinois (15 km west) was 8.2 kg N $ha^{-1}$ $yr^{-1}$. Soils on the site were in the order of mollisols, predominantly Dana silt loam (fine-silty, mixed, superactive, mesic Oxyaquic Argiudolls) with inclusions of Flanagan (fine, smectitic, mesic Aquic Argiudolls) and Blackberry (fine-silty, mixed, superactive, mesic, Oxyaquic Argiudolls) silt loams. Dana and Blackberry silt loams are moderately well drained, while Flanagan silt loams are somewhat poorly drained. Soils were managed using conventional practices for maize-soybean production in the region, consisting of fall tillage after basalt application with cultivation during planting.

The EW experimental trial design at the Energy Farm consisted of four 0.7 ha blocks each containing four 10 × 10 m subplots, with 10 × 16 m buffer zones separating basalt and control sub-plots, and two 3.8 ha large fields overall (**fig. S1**). In a randomized block design, Blue Ridge basalt rock dust was applied (5 kg m$^{-2}$ equivalent 50 t ha$^{-1}$) to two of the subplots within each the four 0.7 ha blocks and one entire 3.8 ha field to create eight 100 m$^2$ treated subplots and one 3.8 ha treated plot, with the other eight 100 m$^2$ subplots and 3.8 ha plot serving as controls. Given that subplots within a block are not true replicates, both subplots of the same treatment within a plot were sampled independently and data was then averaged together for a total experimental design $n$ of 5. All plots were managed in a maize (*Zea mays* L.)-maize-soybean (*Glycine max* L.) rotation typical of the region. Nitrogen fertilizer was applied each year prior to maize planting as 28% urea ammonium nitrate (UAN) at 202 kg N ha$^{-1}$; fertilizer was not applied to the soybean year in the rotation.

Rock dust was applied annually in November 2016, 2017, 2018, and 2019 using conventional lime spreading equipment, with the material subsequently tilled into the soil within 24 hours. Having undergone medium temperature and pressure regional metamorphism, the Blue Ridge basalt is classified as a chlorite-actinolite metabasalt (*17*),(*19*),(*51*). The material for our trials, a by-product of mining from Catoctin Greenstone Belt in Pennsylvania, was purchased from Rock Dust Local LLC, Bridport, Vermont, USA. Blue Ridge basalt mineralogy and chemistry is well characterized (*17*),(*19*),(*51*), and the applied material had a particle size distribution defined by a $p80$ of 267 µm diameter (i.e., 80% of particles are less than or equal to this value).

To investigate whether yield increases with EW and basalt were solely driven by pH increase, we added a limed plot treatment to increase soil pH (Fig. 4). Conventional management in this region generally includes the addition of CaCO$_3$ approximately every five years, with enough lime added to reach a target pH of 6.5. Within each of the four 0.7 ha plots, two additional 10 × 10 m sub plots were established, and lime was added at a rate of 6.725 t ha$^{-1}$ (3 US ton/acre) on April 22, 2020. These plots were sampled and treated statistically the same as the control and basalt sub plots established in 2016, the only difference being the lack of a 3.8 ha limed plot (8 total sub plots, $n = 4$). The granular lime was spread on each sub plot by hand with a pushed broadcast spreader to prevent spread to neighboring plots and cultivated into the soil during planting activities. Measurements were collected as for all other plots.

Aboveground and belowground biomass were collected each year at peak biomass (peak estimated by leaf area index measurements, data not shown) and yield was collected immediately prior to plot scale harvest in all plots. A 0.75 × 0.75m quadrat was used to collect aboveground biomass in all cases. Samples were sorted by hand into major tissue fractions (leaf, stem, reproductive, containing fruit, flower and seed structures) oven dried at 60°C to constant, weighed, homogenized (Retch Wiley Mill, Thomas Scientific, Swedesboro New Jersey, USA), and ground (Geno Grinder 2010, BT&C Lebanon New Jersey, USA), for tissue analyses. Within each quadrat three soil cores, 30 cm depth 5.08 cm diameter, were taken with a slide hammer (AMS Inc., American Falls Idaho, USA), divided into 0-10 and 10-30 cm depth increments and pooled. Soil was removed using an elutriator and roots were oven-dried, weighed, and ground for nutrient analysis. Grain was collected for yield prior to combine harvest for maize and soybean using the same quadrat and was separated by hand from the vegetative fraction, dried, weighed, and ground using the same techniques to determine yield and tissue analysis of removed material.

Soil cores for pH analysis ($n = 8$) were collected at bi-monthly intervals throughout the growing season, and always prior to basalt application for various analyses. Cores were collected with a JMC Backsaver (JMC Soil Samplers, Newton Iowa, USA), 30 cm depth, 3.175 diameter probe, and separated into 0-10 and 10-30 cm increments. Soils were air dried to constant moisture, crushed (DC-5 Dynacrush Soil Crusher, Custom Laboratory Equipment, Holden Missouri, USA), and sieved to pass 2 mm.

### Basalt weathering and carbon dioxide removal rates

**Soil mass balance**. Our approach to track carbon dioxide removal (CDR) rates via changes in the chemical composition of the soil uses cation accounting and detrital, immobile trace element (e.g., titanium, Ti) concentrations to allow for basalt-soil mixture mass balance calculations (*13*). Using immobile elements, we infer a baseline abundance of basalt relative to soil for a given sample. After this determination, a mixing line can be established from the predicted composition of the major base cations sodium (Na), magnesium (Mg), and calcium (Ca) (*13*). The predictions are based on empirically measured major- and trace-element abundances from soil and basalt end-members—samples taken and analyzed prior to spreading basalt on the field. Conceptually, offsets from the mixing line represent loss of basalt or gain of other material associated with the major cations in the soil-basalt mixture within the field. A gain in measured base cations could be the result of soil deposition or the local physical movement of material in a field. Including the offset from the so-called "mixing line", the theoretical mixtures resulting from basalt, soil, and rock weathering may be thought of as a three-component mixing problem with three distinct compositional endmembers: (i) basalt feedstock, (ii) initial soil, and (iii) dissolution (basalt dissolution + soil weathering).

The regions of possible basalt-soil-dissolution combinations can be visualized by plotting an immobile, detrital element on the x-axis and a mobile cation on the y-axis (e.g., titanium and sodium in **fig. S6**). We plot the elemental concentrations as triangular regions that graphically represent the extent of cation loss and carbon capture due to the partial dissolution of basalt added to the soil. The upper left side of the triangular region represents the linear compositional mixture of soil and basalt prior to any cation loss or dissolution and soil weathering. The lower left portion of the triangle represents the maximum amount of cation loss possible from dissolution of basalt and weathering of the soil. The right portion of the triangle is defined by a vertical line with constant values for the composition in the detrital element. This vertical leg of the triangular region represents the assumption that the detrital element is immobile in the mass balance calculations for cations in the soil. Lastly, the vertical leg of the triangle is defined by the cation content for the soil prior to basalt application. The top portion of the dissolution locus for weathering represents cation loss due to basalt dissolution whereas the bottom portion represents additional possible cation loss from soil weathering. The minimum vertex of the triangle represents a soil/basalt mixture for which cation loss proceeds to completion. This set of vertices and horizontal loci is referred to here as the "mixing-triangle."

The $CO_2$ removed from the system during enhanced rock weathering of basalt is due to the conversion of carbonic acid to bicarbonate. This is stoichiometrically proportional to the summation of cation loss over the major base cations of the dissolved basalt. This is a corollary of the Steinour formula, which summarizes the carbon capture potential of chemically weatherable silicate materials (see *13*; and references therein). In the framework presented here, the extent to which the sodium oxides of a given basalt/soil mixture may contribute to carbon sequestration is summarized by the upper portion of the triangular region in **fig. S2**. The leg of the mixing triangle labelled "basalt dissolution" contributes no cations and has the same detrital-element composition as the basalt. Thus, the "dissolution" endmember represents cation loss and the relative buildup of the immobile element in the soil over some unspecified change in time. The proportion of the three mixture-component for a sample within the mixing triangle can be solved algebraically. The linear system is comprised of three equations and three unknowns. The equations themselves describe the bulk composition of a sample as the mass-fraction-weighted average of each component, while the overall the system of equations must enforce mass conservation.

Together, the considerations for bulk composition and conservation of mass are written:

$$C^{\mathcal{D}} = x_s \, c_s^{\mathcal{D}} + x_b \, c_b^{\mathcal{D}} + x_d \, c_d^{\mathcal{D}}, \quad \text{1a}$$
$$C^{+} = x_s \, c_s^{+} + x_b \, c_b^{+} + x_d \, c_d^{+}, \quad \text{1b}$$
$$1 = x_s + x_b + x_d. \quad \text{1c}$$

Here, superscript $\mathcal{D}$ represents detrital and the + denotes a given cation. The $x_{[\cdot]}$ and $c_{[\cdot]}^i$ terms represent the mass fraction and concentration of the soil, basalt and dissolution endmembers which are labelled by *s, b* and *d* subscripts, respectively. The superscript *i* is a placeholder for either the detrital tracer or the cation. The $C^i$ terms represent the bulk composition of the detrital element or cation in the soil-basalt mixture after dissolution (i.e., the blue dot in **fig. S2**).

In previous rock weathering literature, the extent of weathering is typically summarized by a mass transfer function, $\tau$ (see *13*; and references therein). In constructing the mass transfer function, a sub- or superscript, $i$, typically represents an un-weathered parent rock composition while a script $j$ represents the weathered product of the parent rock after some time. In the theoretical framework presented here it is implied that basalt weathering proceeds at a much faster rate than for background soil. Thus, bulk cation loss is encapsulated by the weathering of basalt alone. Relating the framework presented here to a conventional mass transfer function formulation, the parent rock mass fraction contribution for a given cation in basalt is simply $x_b$ while the total mass fraction of each reservoir containing a given cation from the partial weathering of basalt is $x_b + x_d$. The same analysis can be conducted in plots with no basalt amendment to confirm that assumption of negligible natural weathering rates in the uppermost portion of the soil.

**Treatment of baseline soil samples at the Energy Farm.** The principle of our empirical CDR monitoring method — being derived from cation and immobile tracer mass balance — is relatively straightforward. However, for accurate cation loss calculations (and thus accurate tracking of initial carbon dioxide removal rates), baseline values for the soil and basalt compositions, $c_s^i$ and $c_b^i$ must be known. Using our isotope-dilution ICP-MS method, the analytical error in soil measurements is very low (~1%). Although the principle is straightforward and the measurements are precise, the scheme presents logistical sampling complications. Because the composition of the soil must be known, geospatially referenced soil samples should ideally be used as baseline or control values for $c_s^i$. The observed heterogeneity in soil composition over short spatial wavelengths presents a significant difficulty and source of uncertainty. Specifically, for the case of the Energy Farm, samples were segmented into "blocks", "sub-blocks" and "controls".

Despite the segmentation of the study into smaller sub-areas, the natural heterogeneity of soils presents a challenge. At the Energy farm, A, B, C and D denote study sub-blocks and are represented by the red squares in **fig. S3** in order from left to right. In the case of block 3, basalt was applied to sub-blocks A and B while C and D were left as controls. Geospatial information was limited to the sub-block ($10^1$ meter scale). The proximity to soil boundaries (**fig. S3**) in block 3 (Flanagan silt loam (154a) transitions to Dana silt loam (56b) and the spatial variation in baseline soil chemistry renders it challenging to use sub-blocks C and D for tracking temporal compositional changes within sub-blocks A and B. We thus make the distinction between "control blocks" and "baselines." In this context a baseline is a $t = 0$ initial value (i.e., prior to basalt application) taken in the same location as the sample. Control blocks are not directly used for cation loss calculations as the spatial variability of the soils has proven to be too large.

In the case of block 3, plotting sodium, calcium, and magnesium against a ratio of immobile, detrital elements — titanium, Ti, and thorium, Th — for each sample taken at t = 0, we see that sodium varies systematically across the blocks at the beginning of the study (**fig. S4**). Given the resolution of spatial metadata associated with individual samples, we use detrital elemental compositions and sodium as a diagnostic cation for assigning the baseline soil values. Prior to fitting a function to the sodium versus titanium-to-thorium trend, we first checked to make sure that there is a minimal systematic difference in the titanium to thorium ratio throughout the experiment. That is, the addition of basalt consistently adds a similar amount of titanium and thorium annually, as basalt rock sourced from the same parent rock was applied year after year. **Fig. S5** demonstrates that there is little systematic change in the Ti/Th ratio with time. This indicates that using a post-hoc Ti/Th ratio should be appropriate for tethering a basalt + soil mixture to a baseline value. This is not altogether surprising because the application rate for basalt at the Energy Farm experiment remained a constant 5 kg m$^{-1}$ yr$^{-1}$. We use the systematic trend in sodium versus the titanium-to-thorium ratio across all experimental blocks at the energy farm to fit a linear trend. Next, the baseline values and the data points for each block are projected onto the linear trend. Data are connected with their baselines by finding the minimum distance between the trended data point and all potential controls within the block along the linear array. This procedure is repeated block wise so that data points are only linked to baselines within their own block.

We use these controls as baseline values and pair them with non-trended values from after the first year of study (i.e., 2017). We refer to a sample and its appropriate baseline together as a "sample pair". Throughout this study the compositional data from the previous year serves as the baseline value to form sample pairs for mass balance calculations. For the first year of the study, 2016-2017, the pairs selected using trend corrected controls serve as baselines.

**Determination of feedstock weathering in treated soils.** The titanium concentration of the treated samples is scaled by using the expected concentration of titanium after application. The theoretical – or "ideal" – concentration of titanium after basalt amendment is calculated using the application rate ($\mathcal{A}$, kg m$^{-2}$) and basalt density ($\rho_b$, kg m$^{-3}$). Here, we assume that basalt is applied evenly to the field during deployment and is tilled into the underlying soil such that the two solids are well-mixed. Assuming an application rate $\mathcal{A}$ the thickness of an evenly applied basalt layer prior to tilling is given by:

$$h_b = \frac{\mathcal{A}}{\rho_b}. \quad (2)$$

A core sample taken to the depth of tilling should recover the entirety of the basalt application, even after tilling. The portion of the cored sample that is pre-basalt amended soil is therefore,

$$h_s = h_d - h_b, \quad (3)$$

where the depth of the core is given by $h_d$. This statement approximates the expected fraction of background soil in a given sample core (i.e., although the terms $h_{[\cdot]}$ have units of length this argument is volumetric as the cross-sectional area of the core is assumed constant). Therefore, it can be stated $h_s \times A_{\text{core}} = V_s$ and $h_b \times A_{\text{core}} = V_b$. Using $\rho_s$ and $\rho_b$ the mass fraction of soil may be calculated according to:

$$\widetilde{x}_s = \frac{\rho_s V_s}{\rho_s V_s + \rho_b V_b}, \quad \text{and} \quad \widetilde{x}_b = \frac{\rho_b V_b}{\rho_s V_s + \rho_b V_b}. \quad (4)$$

Using Eq. (4), a representative titanium concentration (or any detrital element) can be added to the baseline value for any given application rate and year in the study:

$$[Ti]_i^* = \widetilde{x}_s [Ti]_i^{t_0} + \widetilde{x}_b [Ti]_i^{f} \quad . (5)$$

Here, the subscript $i$ denotes the sample number. The superscript $t_0$ indicates the titanium concentration of the soil prior to basalt application in a given deployment year and f represents the titanium concentration of the basalt feedstock. The superscript $[*]$ indicates the expected value for overall titanium concentration of the sample which is simply the mass-fraction weighted sum of the soil and basalt titanium concentrations.

Using Eq. (5) an expression for the normalized expected titanium content is given as:

$$\Delta^*[Ti]_i = \left( \frac{[Ti]_i^{t_f} - [Ti]_i^{t_0}}{[Ti]_i^* - [Ti]_i^{t_0}} \right) - 1 \quad . (6)$$

Here, the superscript $t_f$ indicates the time when the soil sample was collected after basalt application and weathering. After the first year of the study (2016-2017) treated samples and baselines are paired at random within their corresponding block. This ensures sufficient colocation of the baselines and samples.

We next fit a normal distribution to the Ti, Ca, Mg, and Na data. Because Ti is considered immobile and it is assumed that the application of basalt is even on average at the plot/field scale, the Ti distribution is shifted so that it is centered upon the expected concentration that we predict based on the feedstock Ti content and the feedstock spreading rate. Similar normal distributions are fit to the baseline sample concentrations for Ti, Ca, Mg, and Na.

Next, random values are drawn from the normal distribution the Ti, Ca, Mg, and Na values. The values for Ti, Ca, Mg, and Na together form a single "synthetic" sample. Synthetic samples are not strictly reflections of the composition of the soil at any given point in the field. However, the range and distribution of the synthetic samples are fit from the measured composition of the soil given all valid empirical data from the field. Synthetic samples are coupled with randomly drawn values from the distribution fit to baseline concentrations to form a new synthetic pair of samples (**fig. S7**).

Finally, the now uncertaintiy bounded CDR potential is then calculated using synthetic sample pairs and Equations 1a -1c, where Ti is used as the detrital/immobile element. Synthetic sample pairs are retained for the calculation of the CDR potential only if the Ti content of the treated sample is greater than the baseline. If this condition is not met, this implies that the basalt application resulted in a negative mass addition, which is physically inadmissible. In other words, the full distributions used to generate the synthetic samples allow full exploration of the parameter space but overlap of these distributions may sometimes result in "physically impossible" mixtures that are not considered to when calculating the CDR potential.

For each individual deployment year (and corresponding synthetic sample pair) the solution for the mass fraction of basalt added and dissolved ($x_b$ and $x_d$) can be used to construct a dissolution fraction factor:

$$F_d^+ = \frac{x_d^+}{x_b^+ + x_d^+} \quad . (7)$$

This dissolution factor quantifies the fraction of weathered (or dissolved) minerals bearing the cation [+] ∈ [Mg, Ca, Na] relative to the total added cation abundance. The amount of $CO_2$ taken up per unit area can then be calculated using Eq. (7) and the application rate $\mathcal{A}$. For example, the moles of Mg that contribute to CDR potential may be calculated according to:

$$\Delta M^{Mg} = F_d^{Mg} \times \mathcal{A} \times c_b^{Mg} \times \eta \left(\frac{10^{-3}}{m^{Mg}}\right), (8)$$

where the result of Eq. (8) is in mol m². Here, $\eta$ is a stoichiometric coefficient for the base cation. For Mg and Ca, $\eta = 2$, while for Na, $\eta = 1$. The result for $\Delta M^{Mg}$ can be converted to kg$CO_2$ removed on a per m² area by:

$$\Delta^{Mg} CO_2 = \Delta M^{Mg} \times (M^{CO_2} \times 10^{-3}), (9)$$

where $M^{CO_2} = 44.01$. The same procedure can be repeated for all members of the cation set [+] ∈ [Mg, Ca, Na] for a given sample data point. The sum of the $CO_2$ absorbed gives an approximation to the Steinour formula and is the time-integrated cation loss reported at the point in time where soil sampling took place.

It is important to note that some combinations of data for $C^D$ and $C^+$ may yield a negative value for $\Delta M^+$ for any of the cations considered. This is treated here as secondary mineral precipitation $CO_2$ emission in the full accounting for the CDR potential calculation. Because each cation is considered separately, some calculations will feature a cation balance acting as a $CO_2$ source while the others act to draw down $CO_2$. In other words, $\Delta M^+$ is independent across Mg, Ca, and Na and need not have the same sign across the set of calculations.

**Strong acid weathering correction.** We applied a correction to deduct initial CDR resulting from the dissolution of basalt feedstock by acidity sourced not from carbonic acid. This strong acid correction assumes in-field application rates of urea ammonium nitrate (UAN) fertilizer and measured nitrogen use efficiencies, with the remaining ammonium assumed to be converted to a strong acid input through nitrification. Exchangeable acidity is not incorporated into our initial CDR estimate, given that the exchangeable fractions are assumed to be roughly in equilibrium.

**Initial CDR estimation.** The CDR calculations through the study years are presented in **fig. S8** where two interpretations of the model results from synthetic data pairs are considered. Dots with connecting solid lines represent the mean value for CDR calculations while the shaded regions represent $\pm 1\sigma$. Control plots do not display any resolvable signal for cation loss, and, therefore, we did not do a baseline weathering subtraction. We show two scenarios. In the first (orange region in **fig. S8**), all models with negative values for $\Delta M^+$ are considered to reflect cation addition through net authigenic phase formation (above the mixing line in **fig. S2**) and all sample pairs where the treated cation concentration is lower than its respective baseline are considered fully weathered (below the dotted line in **fig. S2**). This scenario is expected to yield a high estimate of CDR potential in the early years of the trial. In particular, weathering of mineral-bound Ca and Na may be over overestimated due to the natural variation in baseline sample composition compared to the treated sample compositions. The Ca and Na trends in depicted in **fig. S7** illustrate that low cation content in treated samples could lead to an interpretation of strong basalt weathering rates during the study year 2016-2017.

In the second, conservative, scenario (green region in **fig S8**), negative values for $\Delta M^+$ are considered to reflect negligible CDR, as are sample pairs for which the treated cation concentration is lower than the baseline value. The graphical interpretation of this scenario in **fig. S2** is that synthetic data that plot outside of the mixing-triangle are assigned CDR potential of zero in the calculations. This approach filters out the effects of synthetic samples that suggest very high weathering rates.

The average CDR calculation across all experimental blocks and years may be normalized by the total duration of the experiments and the corresponding yearly application rate to obtain an adjusted annual CDR calculation for both scenarios (**fig. S9**). We find that the cumulative distribution of mean CDR calculations is normally distributed. In **fig. S9**, we show an example where randomly selected synthetic data across blocks and years are sampled with replacement. The cumulative average is calculated after each sampling of the data. This process is conducted 100 times, and the individual average annual CDR calculations plotted as semi-translucent dots such that color saturation indicates point density for each scenario. This analysis demonstrates the convergence of average CDR calculations to a mean value despite contributions from models and data that lie well outside the expected mean. This holds true for all presented scenarios. In this analysis, CDR calculations based on single data points may give spurious results whereas the mean value is likely to be representative of results at the field scale.

**Nitrous oxide measurements**
Soil $N_2O$ collections were made at 10-minute intervals over a 30-minute incubation with 3.6 L static chambers atop 20 cm diameter PVC collars inserted approximately 5 cm in the soil, following protocols established for the site (*52*) and previously described (*53*). Collars were kept weeded and free of vegetation at least 24 hours prior to measurement and roots were allowed to grow beneath them. One collar was installed in each subplot of the 0.7 ha plots, and four collars in each 3.8 ha block. Collections began each year prior to planting in the spring and continued throughout the growing season. Sample dates were concentrated around planting and nitrogen fertilization activities in the spring and became less frequent as the season progressed. 15 mL of chamber gas was collected with a syringe and needle inserted through the silicone septa of the chambers and stored in 10 ml glass evacuation vials with 3–mm rubber septa. Samples were analyzed within 1 week by gas chromatography on a Shimadzu GC-2014 gas chromatograph (Shimadzu Scientific Instruments), and fluxes were calculated from measured $N_2O$ concentrations. Fluxes were reported in mg $N_2O$ $m^{-2}$

day$^{-1}$, and cumulative N$_2$O emissions calculated by linear interpolation between field measurements (**fig. S10; Data File 4**) as described in (*52*).

**Plant tissue chemical analysis**
Homogenized plant powders were weighted to 200 mg in tubes followed by sequential addition of 3 ml of ultra-pure concentrated HNO$_3$ (Primar grade, Fisher Scientific, UK), 3 ml Milli-Q H$_2$O, and 3 ml H$_2$O$_2$. Acid digestion was carried out on a microwave (Anton Paar Multiwave) set at the following 3-step program: step 1 (temperature 140°C, time 10 mins, fan level 1), step 2 (temperature hold at 140°C, time 20 mins, fan level 1), and step 3 (temperature 55°C, time 15 minutes, fan level 3). Upon cooling at room temperature, the liquid phase of the resulting tissue digests was filtered through 0.2 µm syringe filter (Sartorius Minisart) and analyzed for multi-elemental composition via ICP-MS (procedure described in detail section 6 below). Tissue concentrations of silicon were measured via pre-calibrated X-ray fluorescence (XRF; Thermo Scientific Niton XL3t GOLDD+) analysis of pelletized plant powders under helium atmosphere (*54*). Dried and ground plant tissues were analyzed for C and N content on an elemental analyzer (Costech EA 4010 CHNSO Analyzer, Costech Analytical Technologies, Valencia CA, USA) using apple leaves, acetanilide (National Institute of Science and Technology, Gaithersburg MD, USA), and certified soil reference material (Leco, St. Joseph MI, USA) as standards.

Obtained tissue concentrations (in mg nutrient kg$^{-1}$ dry biomass) corrected for background matrix effects using blank values were subsequently multiplied with their respective harvested dry biomass weight (in kg dry biomass ha$^{-1}$) to obtain a nutrient pool figure (kg nutrient ha$^{-1}$). Total peak biomass nutrient pool estimates were based on the sum of root, stem, leaf, and floral nutrient pools collected at August peak harvest. Vegetative peak biomass nutrient pool figures were based on the sum of root, stem, and leaf pools for the given nutrient. Total grain biomass pools were derived by multiplying grain tissue concentration with grain dry biomass collected upon the September yield harvest. Replication for each of the two treatment (basalt-treated and control) was $n$ = 10 samples (2 subplots x 4 small sites = 8 and 2 samples from the big plot) per pool per annum to a total of 80 measurements per nutrient pool across the 4-year trial period.

Nitrogen isotope analysis (δ$^{15}$N) of plant samples was undertaken by Iso-Analytical Ltd., Crewe, UK using Elemental Analysis - Isotope Ratio Mass Spectrometry (EA-IRMS). In this technique, samples and references are weighed into tin capsules, sealed, and loaded into an auto-sampler on a Europa Scientific elemental analyser. From there they are dropped in sequence into a furnace held at 1000 °C and combusted in the presence of oxygen. The tin capsules flash-combust, raising the temperature of the sample to ~1700 °C. The gases produced on combustion are swept in a helium stream over combustion catalyst (Cr$_2$O$_3$), copper oxide wires to oxidize hydrocarbons and silver wool to remove sulphur and halides. The resultant gases, N$_2$, NO$_x$, H$_2$O, O$_2$, and CO$_2$ are swept through a reduction stage of pure copper wires held at 600 °C. This step removes O$_2$ and converts NO$_x$ species to N$_2$. A magnesium perchlorate chemical trap is used to remove $H_2O$ and a carbosorb trap is used to remove $CO_2$.

Nitrogen is resolved using a packed column gas chromatograph held at an isothermal temperature of 100 °C. The resultant chromatographic peak for $N_2$ enters the ion source of a Europa Scientific 20-20 IRMS where it is ionized and accelerated. Nitrogen gas species of different mass are separated in a magnetic field then simultaneously measured using a Faraday cup collector array to measure the isotopomers of $N_2$ at m/z 28, 29, and 30. Both references and samples are converted to $N_2$ and analysed using this method. The analysis proceeds in a batch process in which a reference is analyzed followed by a number of samples and then another reference. The reference material used during analysis of your plant samples was IA-R001 (wheat flour, $\delta^{15}N$ AIR = 2.55 ‰).

For quality control purposes check samples of IA-R001, IA-R045 (ammonium sulphate, $\delta^{15}N$ AIR = -4.71 ‰) and IA-R046 (ammonium sulphate, $\delta^{15}N$ AIR = 22.04 ‰) were analysed during batch analysis of your plant samples. IA-R001, IA-R045 and IA-R046 are traceable to IAEA-N-1 (ammonium sulphate, $\delta^{15}N$ AIR = 0.4 ‰). IAEA-N-1 is an inter-laboratory comparison sample distributed by the International Atomic Energy Agency (IAEA), Vienna.

**Soil analyses**
Soil pH was measured in 1:2.5 soil-to-water paste solution following previously described procedures (*55*) with slight modifications. Briefly, to an aliquot of 10 g air-dried soil (sieved <2mm) 25 ml of ultrapure water was added in a 50 ml Falcon tube. Samples were mixed by vortexing at maximum speed for 10s. pH measurements were obtained to the third decimal place using a SciQuip benchtop Precision pH/Ion meter 930 meter calibrated using pre-made buffer solutions of pH 4.00 and pH 7.00.

Soil exchangeable acidity was measured using a soil-buffer equilibration method (*56*) with modifications. We substituted 0.1M sodium phosphate buffer for the highly toxic Shoemaker-McLean-Pratt buffer reagent containing high amounts of chromium (*57*). Briefly, 10 ml of 0.1M sodium phosphate buffer (set at pH 7.00 by using a v/v mixture of 30.5% 0.2M $Na_2HPO_4$, 19.5% 0.2M $NaH_2PO_4$, and 50% ultrapure water) was added to an aliquot of 4 g air-dried soil (sieved <2mm). Samples were briefly mixed by vortexing and incubated at 4°C for 72h. Samples were then inverted several times until fully mixed and pH measured using the benchtop pH meter described above. The difference between pre-equilibration initial buffer pH of 7.00 and post-equilibration pH was used to calculate levels of exchangeable acidity. The difference in pH (ΔpH) was converted to mEq $H^+$ 100 $g^{-1}$ soil by using a standard titration curve of 0.1M sodium phosphate buffer titrated with 0.1M ultra-pure HCl (Normatom®, VWR). The obtained titration curve (**fig. S24**) was linear between pH 6-7 (max. ΔpH = 0.90), covering the range exhibited by our incubated soils (max. ΔpH = 0.68).

Soil exchangeable cations were analyzed by extracting soils with a saturated salt solution following previously described methodology (*55*). Briefly, 20 ml 1M ammonium acetate (pH adjusted to pH 7.00 using ammonium hydroxide and acetic acid) were added to an aliquot of 2 g air-dried soil (sieved <2mm). Samples were shaken on an orbital shaker (Stuart SSL1) at 250 rpm, room temperature for 2 hours. Samples were subsequently centrifuged at 4700 x g for 5 min (Thermo Scientific Heraeus Megafuge 40). Using a 0.2 µm syringe filters, supernatants were additionally filtered into fresh tubes. The resulting filtrates as well as three blanks (ultrapure water) were acidified with ultra-pure nitric acid (Normatom®, VWR) to a total acid concentration of 2% and analyzed via ICP-MS. The mean elemental composition of blanks was subtracted from each of the samples. Cation exchange capacity (CEC) was calculated as the sum of base cations (converted from mg kg$^{-1}$ soil to mEq 100 g$^{-1}$ soil) and exchangeable acidity. Base saturation (sum of exchangeable base cations: $Ca^{2+}$, $Mg^{2+}$, $K^+$, $Na^+$) and exchangeable acidity were subsequently expressed as a percentage of total CEC.

The availability of many soil nutrients depends on soil pH because pH controls the speciation of nutrients in solution and their subsequent adsorption/desorption characteristics, and ultimately their concentration in soil solution and availability to crops (*58*),(*59*),(*60*). Many soil extractants are set at specific pH (e.g. 1M ammonium acetate at pH 7.00). This makes conclusions regarding resulting nutrient concentrations, and particularly their availability to plants difficult. To resolve that, we reconstituted air-dried soil samples with ultrapure water thus bringing the soil sample to its native pH range. Reconstituted soil samples were also briefly heated to boiling temperature to allow for greater desorption that ultimately increases concentrations and improves ICP-MS detection (*61*). In short, 10 g of <2mm-sieved air-dried soil was reconstituted with 25 ml ultrapure water in 50 ml Falcon tubes. The resulting samples were vortexed at maximum speed for 10 seconds and placed in a water bath filled with distilled water pre-heated to boiling temperature. The samples were kept in the boiling hot water bath for 20 minutes, let to cool briefly and centrifuged at 4700 g for 5 minutes. The supernatant was filtered using 0.2 µm syringe filters and placed into fresh tubes. Sample filtrates and at least three blanks were acidified to 2% nitric acid and their multi-elemental composition analyzed using ICP-MS. Sample readings were corrected for background matrix effects by subtracting blank mean values.

*Hot water pH-buffered extractions.* Hot water extraction of soils represent values close to their native range in soil solution. We additionally wanted to estimate the adsorbed pool of Si in our soils irrespective of existing pH differences. Silicon absorption increases with pH and basalt amendment and EW acts to both increase pH thus promote Si adsorption, while it simultaneously releases Si via weathering. Consequently, control samples of lower pH may appear to have similar Si to basalt-treated samples actively releasing Si via weathering because of their higher pH. To overcome this problem, we developed a method that provided means to control for pH and compare basalt-treated and control soils and estimate their soil Si pools at equivalent pH. For this, 10 g of air-dried soils were extracted with 25 ml 0.1M sodium phosphate buffer solutions set up at three different pH values – pH 6.0, pH 6.5, and pH 7.0. The resulting mixtures were left to equilibrate for 72 hours at 4°C. Samples were subsequently extracted using hot water bath as per the operating procedures described in the previous section (hot water extraction). For this test, a sub-sample of 4 samples of basalt-treated and 4 control soils were analyzed from pre-treatment collection ($n = 8$ for 2016) and 5-year post-treatment collection ($n = 8$ for 2021), to a total of 16 samples.

Unless otherwise stated, replication for all other soil tests was as follows: each of the two treatment (basalt-treated and control) was represented by $n = 10$ samples (2 subplots × 4 small sites = 8 and 2 samples from the big plot) × 2 depths (0 - 10cm and 10 - 30cm) × 5 study years (2016-2020).

Air-dried and ground soil samples were analyzed for C and N content on an elemental analyzer (Costech EA 4010 CHNSO Analyzer, Costech Analytical Technologies, Valencia CA, USA) using apple leaves, acetanilide (National Institute of Science and Technology, Gaithersburg MD, USA), and certified soil reference material (Leco, St. Joseph MI, USA) as standards. To obtain nitrogen mineralization rates we calculated the loss of soil N in the top 30 cm for each treatment at each block between 2016 and 2020 (**fig. S12**). These were averaged for treatment and converted to kg N loss ha$^{-1}$ yr$^{-1}$ by multiplying the soil N loss (in %) with estimated amount of soil in top 30 cm and dividing by the number of years. Treatment-specific nitrogen mineralization rates were used in the calculation of nitrogen use efficiency (NUE) and assumed to be constant throughout the 4-year trial period. NUE were calculated by the established methods (*62*) following formula:

$$\text{NUE\%} = \frac{N_{grain} - N_{mineralization}}{N_{fertilizer}} 100$$

where $N_{grain}$ is the total grain biomass N pool (in kg ha$^{-1}$), $N_{mineralization}$ is the treatment-specific nitrogen mineralization rate and $N_{fertilizer}$ is the N fertilizer application rate of 202 kg N ha$^{-1}$ applied as 28% UAN.

Soil samples were prepared for metal analysis using isotope dilution (ID-) ICP-MS at the Yale Metal Geochemistry Center (see Section 6.2). 100 mg of <2mm-sieved and air-dried sample was subjected to leaching with a 1N ammonium acetate solution at pH = 7 to remove the exchangeable fraction from the solid phase (after *2*). The resultant solid sample was dried at 60°C overnight, then heated at 600°C for 24 hours to remove organic matter in the soil mixture. The ashed sample was then dissolved in a mixture of aqua regia (hydrochloric and nitric acid) and hydrofluoric acid and hydrochloric acid as follows: 10ml aqua regia was added to Teflon beakers containing ashed sample and left to sit for 4 hours after which 1ml hydrofluoric acid was added, the beakers capped and left on a hotplate for 24 hours at 100°C. Along with aqua regia and hydrofluoric acid, a 7.5μl aliquot of isotope spike 'cocktail' solution was added at this stage to ensure sample-spike equilibration for ID-ICP-MS (see Section 6.2). Samples were then uncapped and left to dry down on a hotplate at 90°C, after which they were raised in 4ml 6N hydrochloric acid mother solution for storage. 15μl aliquots of mother solution for each sample were dried down in Teflon beakers, then raised in 1% nitric acid to run on ICP-MS.

**ICP-MS procedures and calibration**

*Solution chemistry.* Multi-element analysis of diluted solutions was undertaken by inductively coupled plasma mass spectrometry (ICP-MS) (Thermo-Fisher Scientific iCAP-Q; Thermo Fisher Scientific, Bremen, Germany). Samples are introduced (flow rate 1.2 mL min-1) from an autosampler (Cetac ASX-520) incorporating an ASXpress™ rapid uptake module through a perfluoroalkoxy (PFA) Microflow PFA-ST nebuliser (Thermo Fisher Scientific, Bremen, Germany). Sample processing is undertaken using Qtegra™ software (Thermo-Fisher Scientific) utilizing external cross-calibration between pulse-counting and analogue detector modes when required. The instruments are run employing several operational modes. The iCAP-Q employs in-sample switching between two modes using a collision cell (i) charged with He gas with kinetic energy discrimination (KED) to remove polyatomic interferences and (ii) using $H_2$ gas as the cell gas. Typically, in-sample switching is used to measure Se in $H_2$-cell mode and all other elements in He-cell mode. Peak dwell times are 100 milliseconds for most elements with 150 scans per sample. Internal standards, used to correct for instrumental drift, are introduced to the sample stream on a separate line (equal flow rate) via the ASXpress unit or are added directly to calibration standards and samples and introduced on a single line. Internal standards typically include combinations of Sc (10 µg $L^{-1}$), Ge (10 µg $L^{-1}$), Rh (5 µg $L^{-1}$), Re (5 µg $L^{-1}$), and Ir (5 µg $L^{-1}$).

The matrices used for internal standards, calibration standards and sample diluents are typically either 2% Primar grade $HNO_3$ (Fisher Scientific, UK) with 4% methanol (to enhance ionization of some elements). Calibration standards typically include (i) a multi-element solution with Ag, Al, As, Ba, Be, Cd, Ca, Co, Cr, Cs, Cu, Fe, K, Li, Mg, Mn, Mo, Na, Ni, P, Pb, Rb, S, Se, Sr, Ti, Tl, U, V and Zn, in the range 0 – 100 µg $L^{-1}$ (0, 20, 40, 100 µg $L^{-1}$) (Claritas-PPT grade CLMS-2 from SPEX Certiprep Inc., Metuchen, NJ, USA); (ii) a bespoke external multi-element calibration solution (PlasmaCAL, SCP Science, France) with Ca, Mg, Na and K in the range 0-30 mg $L^{-1}$ and (iii) a mixed phosphorus, boron and sulphur standard made in-house from salt solutions ($KH_2PO_4$, $K_2SO_4$ and $H_3BO_3$). See Data File S6 for limits of blank value expressed in concentration units (BEC) and limits of detection (LOD) for School of Chemistry, Analytical and Scientific Services Nottingham University. Concentrations were in calibration range listed, as expressed in units of mg $L^{-1}$ and µg $L^{-1}$. These were subsequently converted to mg $kg^{-1}$ or µg $kg^{-1}$ soil using extractant volume and extracted soil weight.

*Solid phase soil chemistry.* Multi-element analysis of soil samples (see Section 5.6 for preparation) was undertaken by ID-ICP-MS at the Yale Geochemistry Center. Isotope dilution ICP-MS was used to reduce analytical error when measuring concentrations of Ca, Mg, and Ti (see REF 4). Isotope dilution is an analytical method whereby the concentration of an element in a sample can be measured from the known concentration of an element in a spike solution, and the ratios of two isotopes of the same element in the natural sample and the spike respectively (*63*),(*64*),(*65*). We use an isotope spike 'cocktail', doped with isotopes of Mg, Ti, and Ca found in lower abundance in natural samples. Isotope spikes were prepared from powders of spiked $TiO_2$, $MgCO_2$ and $CaCO_2$. The pure spike Mg and Ca carbonate powders were digested using HCl, and the Ti oxide using $HNO_3$+HCl+HF. Following the digest each spike solution was calibrated by measuring the relative concentration of Mg, Ca, and Ti isotopes on a Thermo-Fisher Scientific Neptune Plus multicollector ICP-MS (Thermo-Fisher Scientific, Bremen, Germany) for ~48 hours (13). Estimate of the error on the spike determination are < 0.1 ‰ based on replicate analysis. Individual spike solutions are then used to make an isotope spike 'cocktail' solution containing Mg, Ca and Ti spikes. To run soil samples prepared for ID-ICP-MS, two models of ICP-MS were used: a Thermo-Fisher Scientific Element High Resolution Magnetic Sector ICP-MS (Thermo-Fisher Scientific, Bremen, Germany), and a PerkinElmer NexION 5000 Multi-Quadropole ICP-MS (PerkinElmer, Waltham, MA, U.S.A). We employed a bracketing scheme of geochemical reference standards (BHVO-2 and SGR-1B) before and after every block of 5 samples in order to correct for machine drift. Analytical error was determined by using additional reference standards within runs and was comparable for both machines.

**Magnetic extraction of rock grains and rock XRF analyses.** Basalt grains were recovered from basalt-treated field soil samples using sequential magnetic extraction (**fig. S17**). Magnetic extraction of basalt from soils is possible due to the frequent inclusion of magnetic minerals (magnetite, maghemite) in basalts and their relatively rare occurrence in many soils. Preliminary tests revealed that over 30% of the studied meta-basalt particles were magnetically attracted relative to <0.2% of soil particles. Briefly, for field weathered samples ~25 g of air-dried and ground basalt-treated soil samples (0-10 cm collected in August 2020) were placed in plastic weighing boats and strong neodymium magnets tightly covered in thin microscope lens paper wipe was run over the sample for 30 seconds. Collected basalt grains were placed into a fresh weighing boat by releasing the magnet from the paper material over the fresh weighing boat. The procedure was repeated 3 times.

The collected grains were further magnetically separated into another weighing boat by this time only hovering the magnet to select the most magnetic of particles. That was repeated a total of two times before the final selection of basalt grains was placed in an Eppendorf tube. As a baseline, we used fresh unweathered basalt mixed with control soil samples that were brought to field moisture, dried and also separated magnetically as described to account for bias. Magnet-extracted weathered basalt grains ($n$ = 15) and magnet-extracted fresh unweathered basalt grains ($n$ = 5) were scanned using a Thermo Scientific XLt2 GOLDD XRF mounted on a safety stand and remotely operated (66). Samples were placed in a specially tailored cup holder for small samples with the bottom of the holder covered in a transparent Mylar® polyester film for minimal interference during XRF scanning. XRF was specifically calibrated for basalt silicate matrix using an international standard basalt sample BHVO-2. Each grain sample was scanned in Cu/Zn mining mode three times and final values obtained by averaging across the three measurements. Reproducibility for the measured elements was assessed as the standard deviation between three separate measurements of the same sample and it was as follows: Ti (± 1115 ppm), K (± 567 ppm), and P (±1 97 ppm) which represented 6.8%, 7.8%, and 10.6%, respectively, of mean values for these elements in the analysed samples.

**Root RNA extraction and transcriptome analyses.** Healthy green plants of average size were selected in the field from each treatment from different blocks. Roots from crop plants were collected from comparable developmental stages: soybean plants roots were collected at the end of July 2019, at the R4-R5 reproductive stage and roots collected from maize plants mid-August 2020, at the R3-R5 reproductive stage. Each plant was uprooted by gently inserting a foot shovel from each side until loose. Next, plants were uplifted exposing the root ball which was cleaned in a sterilized pan with distilled water to remove contaminating soil particles. Several root fragments were cut into a labelled aluminium foil using sterilized scissors. Aluminium foil packets were immediately snap-frozen in a dry shipper filled to saturation with liquid $N_2$. The entire procedure took no more than 60-90 seconds per plant as to prevent possible RNA degradation. Upon return to the laboratory, samples were placed in a -80ºC freezer prior extraction. RNA was extracted from roots using the RNeasy® Plant Mini Kit (Qiagen) following manufacturer's instructions therein. RNA samples were further purified to remove any contaminating DNA by using a RNase-Free DNase Set (Qiagen). The reaction is subsequently cleaned up by loading the samples onto RNA Clean and Concentrator kit columns (Zymo) with final eluate volume of 50 µl.

The concentration of RNA in samples was checked on a Nanodrop instrument with samples showing an average concentration of 46 ng/µl and a total RNA content of ~2.3 µg. RNA integrity was assessed by gel electrophoresis with all samples exhibiting the characteristic 25S rRNA and 18S rRNA bands (**fig. S20**). To capture the coding transcriptome, cleaned total RNA samples were enriched for mRNA by the polyA tail-selection method using the Kapa RNA HyperPrep kit (Roche). mRNA library preparation was carried out with the TruSeq Stranded mRNA kit (Illumina). Soybean root mRNA samples ($n$ = 15) were sequenced on one NovaSeq S1 lane (Illumina) using 2x150nt paired-read chemistry. Maize root mRNA samples ($n$ = 32) were sequenced on a S4 lane of the NovaSeq equipment using 2 × 150 bp paired-read chemistry.

Sequencing was carried out at the Roy J. Carver Biotechnology Center, University of Illinois at Urbana-Champaign, USA. Preliminary bioinformatics including demultiplexing and removal of adapters was performed by the sequencing facility. Sequencing reads were then uploaded onto the Galaxy Europe (https://usegalaxy.eu/) server (*67*). Paired-read libraries were then further trimmed based on length and quality using the sliding window trimming mode in Trimmomatic with minimum length of 75 bp. The cleaned paired-read libraries were then aligned against their reference genome sequence (Gmax JGI Wm82.a2.v1 for soybean, and Zmays 493 APGv4 for maize) using HISAT2 (*68*). Unaligned reads were discarded and aligned reads were assembled using StringTie (*69*) with average read length of 150 bp and minimum assembled transcript length of 200 bp. Gene counts were then normalized by using factor analysis of control samples via the RUVseq tool (*70*) to normalize for variance originating from samples drawn from different field plots.

Finally, samples were submitted for differential expression analyses through DESeq2 (*71*) using pre-filtering of 1 read per sample and two factors – the primary factor treatment (basalt/control) and secondary factor – block from which samples are derived (**Data File 6**). Furthermore, the RUVseq output (batch effects control method k2) was utilized as an additional batch factor into DESeq2 analyses. DESeq2 statistics for the primary factor treatment (basalt/control) were used to assign genes as differentially expressed when $P < 0.05$. Detailed gene annotation for the particular reference genome version including deflines, KEGG, KO, KOG, PANTHER, and GO functional predictions as well as best-hit-Arabidopsis orthologues were obtained from Phytozome 13 (*72*) (https://phytozome-next.jgi.doe.gov/).

**Economic price of yield increases calculations**
Based on USDA data, we assume the corn belt is 50% corn and 50% soybean, which is broadly equivalent to USA national average; 81 million acres corn / 87 million soybean (*73*). We used a total Corn Belt area of 60 Mha giving 30 Mha (i.e., 74 million acres) for corn and soybean each (*74*). We used 2022 base yields of 173.3 bushel/acre (corn) and 49.5 bushel/acre (soybean) (*75*). Current (Feb. 2023) prices for corn: $6.44 / bushel and soybean: $15.12 /bushel (*76*). Over the course of the last 12 months corn price varied by +26/-3% of current value, and soybean by 11/-12 %.

We calculate the value of increased corn/soy production with EW (Fig. 4) as corn/soybean area × corn/soybean base yield per area × fractional change in corn yield, e.g., for 8.5% yield increase, fractional increase in corn yield would be 0.085. We then calculate value of extra production as: increased yield with EW × current price of corn/soybean. Assuming Feb. 2023 prices, this approach gives:

Corn: 8.5% yield increase with EW —> $ 7.03 bn (±1.2)
Corn: 13.3% yield increase with EW —> $ 11.00 bn (+2.86/-0.33)
Soy: 18% yield increase —> $ 9.99 bn (±1.19/-1.1)
Uncertainty shows effect of using max and min crop value for last 12 months.

**Acknowledgments**
We thank Jason Clark for assistance with sourcing basalt feedstock and Saul Vazquez Reina (Nottingham University School of Chemistry Analytical and Scientific Services) for ICP-MS analyses. Construction of the RNAseq libraries and sequencing on the Illumina NovaSeq 6000 were performed at the Roy J. Carver Biotechnology Center at the University of Illinois at Urbana-Champaign.

**Supporting Information**

Figs S1 to S24
Data files:
Date File 1: Compositional data for Blue Ridge metabasalt feedstock
Date File 2: Compositional data for baseline soil samples (prior to basalt application)
Data File 3: Compositional data for soil samples taken at the Energy Farm EW trials
Data File 4: Energy EW trial soil $N_2O$ fluxes, 2017-2020
Data File 5: Muti-element ICP-MS limits of detection.
Date File 6: Full gene listing and sample counts

**Figure legends**

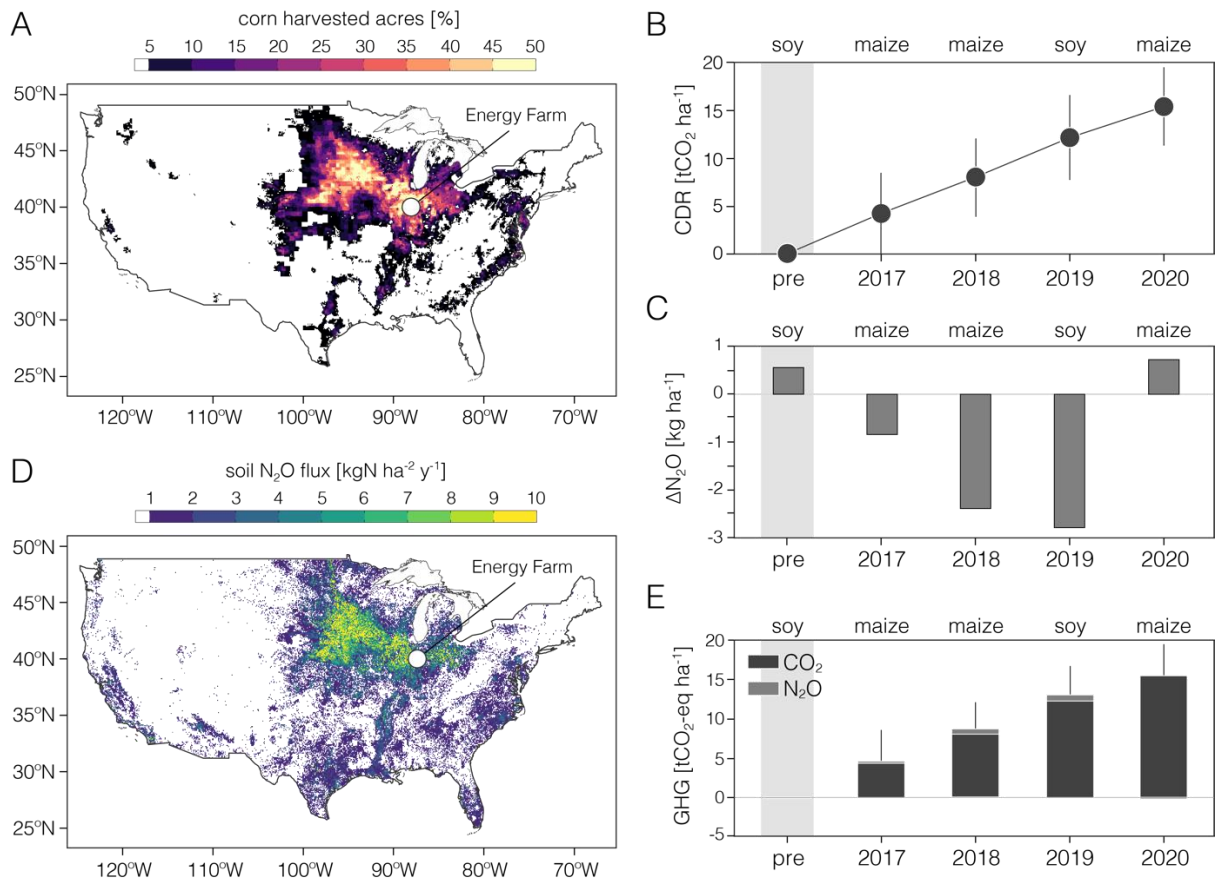

**Figure 1. Carbon dioxide removal and trace greenhouse gas mitigation by enhanced weathering with basalt under field conditions.** (**A**) Harvested corn acres across the U.S. Corn Belt, shown as regional percent area (*40*). (**B**) Measured cumulative carbon dioxide removal (CDR) over our 4-year trial in a Midwestern corn-soy rotation. Results show mean CDR across four sample blocks, with error bars showing ±1SE. Grey bar shows pre-treatment period (pre). (**C**) Cumulative change in $N_2O$ emissions during 4-year field trial relative to control fields. Note that $N_2O$ fluxes in control plots were higher than those of basalt-treated plots prior to basalt spreading. (**D**) Soil nitrous oxide ($N_2O$) fluxes in the coterminous U.S. (*41*). Long-term field site shown with open circle in (**A**) and (**D**). (**E**) Cumulative greenhouse gas (GHG) mitigation through the course of long-term field trial. Fluxes of $CO_2$ equivalent ($CO_2$-eq) are calculated for $N_2O$ assuming a 20-year global warming potential (GWP) for $N_2O$ of 273 (*18*).

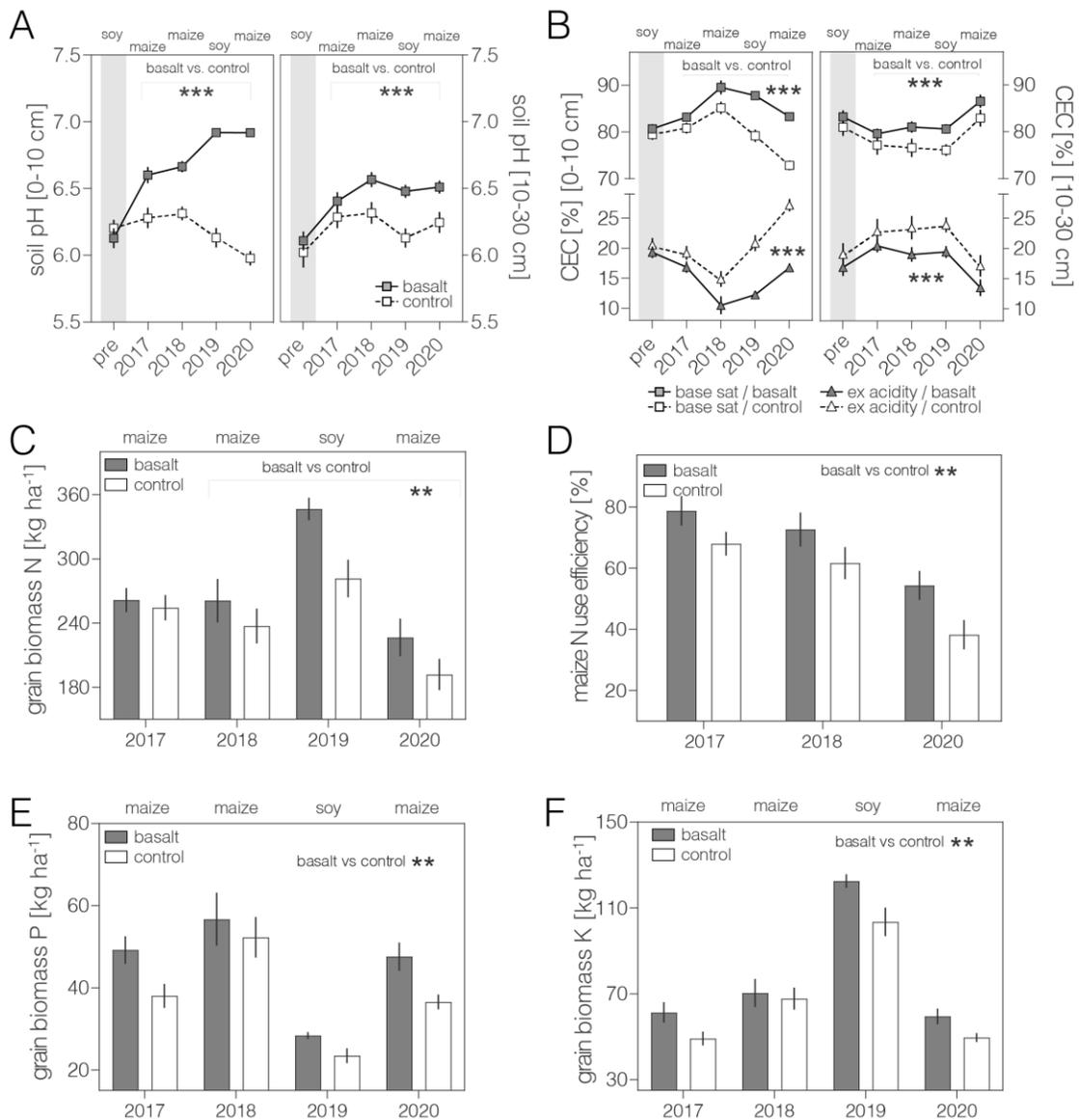

**Figure 2. Soil and crop biogeochemical responses to enhanced weathering under field conditions.** (**A**) Soil pH increased significantly with enhanced weathering (EW) with basalt at 0-10 cm and 10-30 cm depths. (**B**) Soil cation exchange capacity increases with EW while exchangeable acidity decreases at 0-10 cm and 10-30 cm depths. (**C**) Total grain biomass nitrogen (N), (**D**) plant nitrogen-use efficiency (NUE), (**E**) Total grain phosphorus (P) and (**F**) potassium (K) all increased significantly with EW. A to F, Two-way ANOVA basalt vs. control main factor, asterisks indicate significant difference (**$P < 0.05$, ***$P < 0.001$).

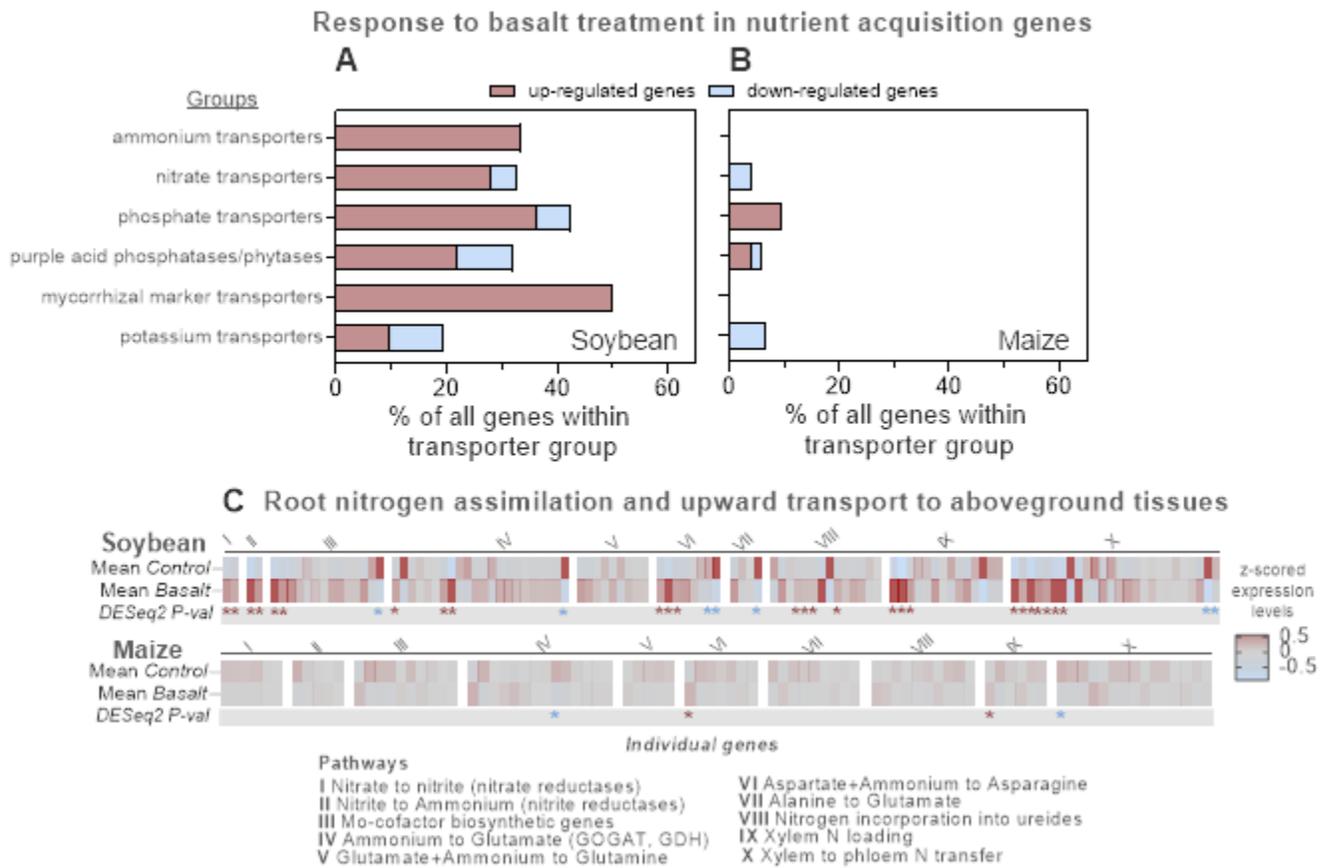

**Figure 3. Root transcriptional responses of crops to enhanced weathering under field conditions.** (**A**) Soybean and (**B**) maize root nutrient transporter and acid phosphatase transcriptional responses to EW. Responses are expressed as percentage of significantly up- and down-regulated transporter genes (differential expression DESeq2 tests, $P < 0.05$) within all genes in a transporter group in response to EW (i.e., between basalt and control plants); (**C**) Heatmap of z-score normalized root expression levels of genes involved in different pathways (labelled I to X) of root nitrogen assimilatory metabolism and long-distance upward transport. The heatmap shows that a whole suite of genes is differentially expressed (DESeq2, $P < 0.05$; blue asterisks denote down-regulated and red asterisks – up-regulated genes) in response to basalt in soybean, but not maize. Results are based on analysis of 47 root transcriptome libraries ($n = 8$ basalt and $n = 7$ control soybean, reproductive R4-R5 growth stage, summer 2019) and maize ($n = 16$ for control and basalt maize, reproductive R3-R5 growth stage, summer 2020). See Supplementary Materials for gene listing in all groups.

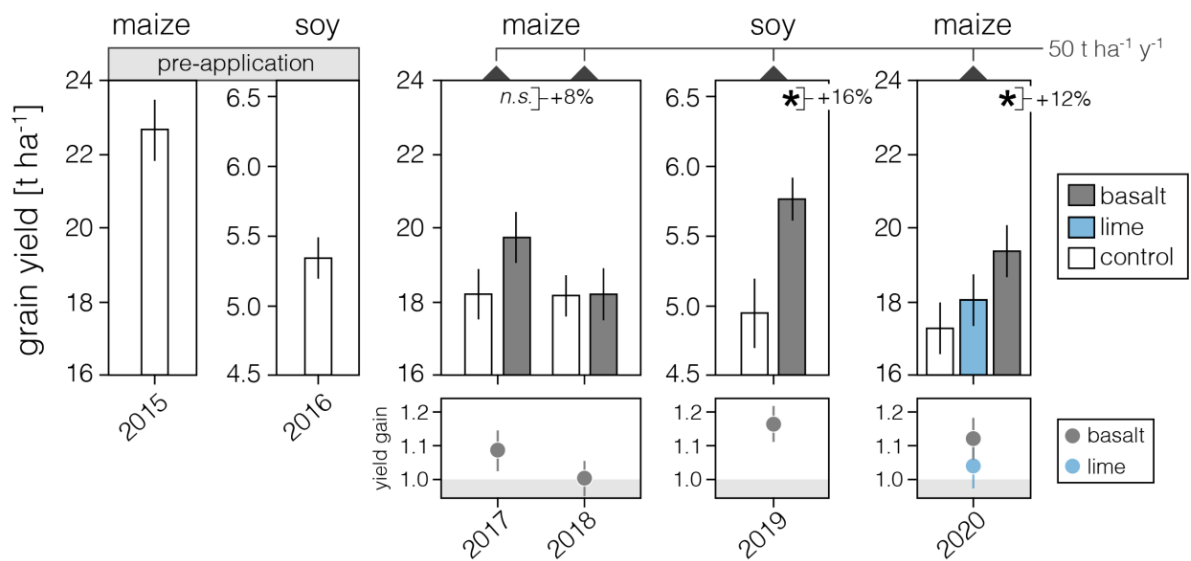

**Figure 4. Grain yield responses of food and energy crops to enhanced weathering under field conditions.** (**A**) Maize and (**B**) soybean. Two-way ANOVA basalt vs. control main factor, asterisks indicate significant difference (*$P$ < 0.05).